\documentclass[aps,pra,onecolumn,showpacs,amsmath,amsfonts,amssymb,superscriptaddress]{revtex4}
\usepackage{xspace,epsfig,amsfonts,amssymb}

\begin{document}

\newcommand{\beq}{\begin{equation}}
\newcommand{\eeq}{\end{equation}}
\newcommand{\beqa}{\begin{eqnarray}}
\newcommand{\eeqa}{\end{eqnarray}}
\newcommand{\ket} [1] {\vert #1 \rangle}
\newcommand{\bra} [1] {\langle #1 \vert}
\newcommand{\braket}[2]{\langle #1 | #2 \rangle}
\newcommand{\proj}[1]{\ket{#1}\bra{#1}}
\newcommand{\mean}[1]{\langle #1 \rangle}
\newcommand{\opnorm}[1]{|\!|\!|#1|\!|\!|_2}
\newcommand{\half}{\smallfrac{1}{2}}
\newtheorem{lemma}{Lemma}

\newcommand{\Tr}{\mathop{\mathrm{Tr}}\nolimits}

\title{A Lossy Bosonic Quantum Channel with Non-Markovian Memory}

\author{Oleg V. Pilyavets} 
\email{pilyavets@gmail.com}

\affiliation{Dipartimento di Fisica, Universit\`{a} di Camerino, 
I-62032 Camerino, Italy}
\affiliation{P. N. Lebedev Physical Institute, Leninskii Prospect 53, 
Moscow 119991, Russia}

\author{Vadim G. Zborovskii}
\email{vzborovsky@gmail.com}

\affiliation{Troitsk Institute for Innovation and Fusion Research,
Troitsk 142190, Russia} 

\author{Stefano Mancini}
\email{stefano.mancini@unicam.it}

\affiliation{Dipartimento di Fisica, Universit\`{a} di Camerino, 
I-62032 Camerino, Italy}

\begin{abstract} 
We provide a simple and realistic model to study memory effects in a lossy bosonic quantum channel over
arbitrary number of uses. The noise correlation among different uses is introduced by contiguous modes
interactions which results in an exponential decay of the correlations over channel uses (modes).  This
model allows us to characterize the asymptotic behavior of the channel for classical information
transmission.
\end{abstract}


\maketitle

\section{Introduction}

A memoryless quantum communication channel makes the fundamental assumption that the noise between
consecutive uses of the channel is independent. This assumption is reasonable for many real-world
applications, but for many others the noise may be strongly correlated between uses of the channel.
Recently a lot of efforts have been dedicated to the development of quantum models that encompass memory
effects (see \cite{Kre} for an overview). The main motivation that has led to investigate such effects in
quantum channels has been the possibility to enhance their classical capacity by means of entangled
inputs.  Such a possibility has been recently put forward in channels with continuous alphabet \cite{GM,
Rug1, Cerf, Rug2, Rug3}. These make use of bosonic field modes whose phase space quadratures enable for
continuous variable encoding/decoding \cite{Bra}.  However, most of these works were limited to two
channel uses in order to provide a proof of principle of the behavior of quantum memory channels.  Since
the notion of capacity is intimately related with the asymptotic behavior of a channel, there is a
persistent wish to move on from small to large (towards infinite) number of channel uses.

The lossy bosonic channel, which consists of a collection of bosonic modes that lose energy en route from the
transmitter to the receiver, belongs to the class of Gaussian channels which provide a fertile testing
ground for the general theory of quantum channels' capacities \cite{Hol} and are easy to implement
experimentally \cite{Bra}.
Here we present  a model for a lossy bosonic quantum channel with memory that can be employed for an arbitrary number of channel uses.  The memory
effects are realized by considering quantum correlations among environments acting on different channel
uses \cite{GM}.  This model allows us to characterize the asymptotic behavior of the channel for
classical information transmission. We then show the usefulness of entangled inputs in presence of
memory. In particular, we show that they can enhance the classical capacity above that of the
corresponding memoryless channel.  We also show the utility of entangled inputs for rates achievable by
conventional decoding procedures (heterodyne and homodyne measurements), although these do not exceed the
classical capacity of the corresponding memoryless channel.  
 
The paper is organized as follows. In Section II we briefly review the lossy bosonic channels.  In Section
III we introduce the model for the memory effects. Then, Section IV is devoted to the study of the
Holevo-$\chi$ quantity, while Section V concerns rates achievable with conventional decoding procedures.
Finally, Section VI is for conclusions.


\section{Lossy bosonic channels}

\begin{figure}
\begin{center}
\includegraphics[width=0.4\textwidth]{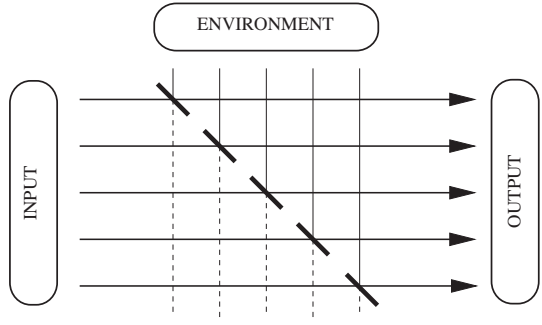}
\end{center}
\vspace{-0.5cm}
\caption{\label{fig_model} 
The model for a lossy bosonic channel $\mathcal{L}$.
Each input mode (left-right line), representing one use of the channel, 
 interacts with the corresponding environment mode (top-bottom line)  through a beam-splitter.
To introduce memory effects, environment modes are initially considered in a correlated state.}
\end{figure}

The model for a lossy bosonic channel $\mathcal{L}$ is depicted in Fig.\ref{fig_model}. 
On each use of the channel we have an input mode. Let us consider $n$ of such modes (uses) with corresponding environment modes. They interact through beam splitters with transmittivity $\eta$. 
If the modes are described by canonical operators $q_j,p_j$, satisfying canonical commutation relation $[q_j,p_k]=i\delta_{jk}$, then the interaction leads to the following transformations \cite{Bra}
\begin{equation}
\begin{split}
&q_j^{in}\,\rightarrow\,\sqrt{\eta}\,q_j^{in}\,+\sqrt{1-\eta}\,q_j^{env},\\
&p_j^{in}\,\rightarrow\,\sqrt{\eta}\,p_j^{in}+\sqrt{1-\eta}\,p_j^{env},\\
&q_j^{env}\,\rightarrow\,-\sqrt{1-\eta}\,q_j^{in}\,+\sqrt{\eta}\,q_j^{env},\\
&p_j^{env}\,\rightarrow\,-\sqrt{1-\eta}\,p_j^{in}+\sqrt{\eta}\,p_j^{env}.
\end{split}
\label{BSn}
\end{equation}

For a memoryless channel,
the environment modes are initially in an uncorrelated state, the vacuum state in the simplest case.
More generally, we can consider a Gaussian state characterized by covariance matrix $V_{env}$ and zero displacement vector. 
In terms of Wigner function we have
\begin{equation*}
W_{env}(\mathbf{x}_{env})=
\frac1{\sqrt{|V_{env}|}}
\exp\left[-\frac12
\mathbf{x}_{env}^{\top}
V_{env}^{-1}
\mathbf{x}_{env}
\right],
\label{Wenv}
\end{equation*}
where  $\mathbf{x}_{env}:=(q_1^{env},\dots,q_n^{env},p_1^{env},\dots,p_n^{env})^{\top}$
with ${}^{\top}$ denoting the transpose and  
$|\bullet |$ the determinant (when applied to matrices).
Throughout the paper the normalization of a $n$-mode Wigner function reads
$\int W(\mathbf{x})\,d\mathbf{x}=(2\pi)^n$.

Let us now consider the classical use of the bosonic channel where
input quantum states carry the values of a random classical variable. 
Then, the mapping $\rho\mapsto\mathcal{L}[\rho]$ can be seen as a mapping of phase space points,
so that for each mode we have two real random values (or one complex) carried.
Therefore, we label an input state over $n$ modes (channel's uses) as
 $\rho_{in}^{\mbox{\boldmath$\alpha$}}$ with  $\mbox{\boldmath$\alpha$}
:=(\Re(\alpha_1),\dots,\Re(\alpha_n),\Im(\alpha_1),\dots,\Im(\alpha_n))^{\top}$ and $\alpha_j\in\mathbb{C}$.

Restricting our attention to input Gaussian states \cite{Hol},
we consider the classical variable $\mbox{\boldmath$\alpha$}$ encoded via random displacements of a suitable (Gaussian) seed state $\sigma_{in}$,  that is
\begin{equation*}
\rho^{\mbox{\boldmath$\alpha$}}_{in}= \left[\otimes_{j=1}^nD_j(\alpha_j)\right]\,\sigma_{in}\,\left[\otimes_{j=1}^nD_j(\alpha_j)\right]^{\dag},
\label{sigin}
\end{equation*}
where $D_j$ denotes the displacement operator on the $j$-th mode and ${\mbox{\boldmath$\alpha$}}$
is chosen according to a Gaussian distribution 
\begin{equation}
P(\mbox{\boldmath$\alpha$})=
\frac{1}{\pi^n\sqrt{|V_{cl}|}}
\exp\left[-
\mbox{\boldmath$\alpha$}^{\top} V_{cl}^{-1} \mbox{\boldmath$\alpha$}
\right],
\label{Pal}
\end{equation}
having classical covariance matrix $V_{cl}/2$.

In the memoryless case, the classical capacity is achieved by using 
$\sigma_{in}=|0\rangle\langle 0|^{\otimes n}$
 and $V_{cl}={\rm diag}(N)$,
where $N$ represents the average number of input photons per mode \cite{GG}. 
In such a case entangled inputs turn out to not be useful.

Quite generally we shall consider $\sigma_{in}$ characterized by a covariance matrix $V_{in}$ and $V_{cl}$ not necessarily be diagonal. Then,
the state $\rho^{\mbox{\boldmath$\alpha$}}_{in}$ will be characterized by the Wigner function
\begin{equation*}
W^{\mbox{\boldmath$\alpha$}}_{in}(\mathbf{x}_{in})=
\frac1{\sqrt{|V_{in}|}}
\exp\left[-\frac12
\left(\mathbf{x}_{in}-\sqrt{2}\mbox{\boldmath$\alpha$}\right)^{\top}
V_{in}^{-1}
\left(\mathbf{x}_{in}-\sqrt{2}\mbox{\boldmath$\alpha$}\right)
\right],
\label{Win}
\end{equation*}
where 
$\mathbf{x}_{in}:=(q_1^{in},\dots,q_n^{in},p_1^{in},\dots,p_n^{in})^{\top}$.
In turn, the average input state, $\overline{\rho}_{in}=\int d\mbox{\boldmath$\alpha$}\ P( \mbox{\boldmath$\alpha$})\rho^{\mbox{\boldmath$\alpha$}}_{in}$,
will be characterized by the Wigner function
\begin{equation*}
\overline{W}(\mathbf{x}_{in})=
\int d\mbox{\boldmath$\alpha$}\ P( \mbox{\boldmath$\alpha$})
W_{in}^{\mbox{\boldmath$\alpha$}}(\mathbf{x}_{in}),
\label{Winave}
\end{equation*}
having covariance matrix
\begin{equation*}
\overline{V}_{in}=V_{in}+V_{cl}.
\label{Vinave}
\end{equation*}

The output state ${\rho}^{\mbox{\boldmath$\alpha$}}_{out}=\mathcal{L}\left[ \rho^{\mbox{\boldmath$\alpha$}}_{in} \right]$
is still Gaussian and it has covariance matrix
\begin{equation}
V_{out}=\eta V_{in}+(1-\eta) V_{env},
\label{Vout}
\end{equation}
as consequence of transformations~(\ref{BSn}).
In fact, the Wigner function of the input and environment is given by
\begin{equation}
W_{tot}^{\mbox{\boldmath$\alpha$}_{tot}}(\mathbf{x}_{tot})=
\frac1{\sqrt{|V_{tot}|}}
\exp\left[-\frac12
\left(\mathbf{x}_{tot}-\sqrt{2}\mbox{\boldmath$\alpha$}_{tot}\right)^{\top}
V_{tot}^{-1}
\left(\mathbf{x}_{tot}-\sqrt{2}\mbox{\boldmath$\alpha$}_{tot}\right)
\right],
\label{Wtot}
\end{equation}
with block covariance matrix
\begin{equation*}
V_{tot}:=
\begin{pmatrix}
V_{in}&0\\
0&V_{env}
\end{pmatrix},
\label{Vtot}
\end{equation*}
and vectors $\mathbf{x}_{tot}^{\top}:=(\mathbf{x}_{in}^{\top},\mathbf{x}_{env}^{\top})$,
$\mbox{\boldmath$\alpha$}_{tot}^{\top}:=(\mbox{\boldmath$\alpha$}^{\top}, 0^{\top})$.
By applying the beam splitter unitary transformation
$\mathbf{x}_{tot}\rightarrow B\mathbf{x}_{tot}$ 
in Eq.(\ref{Wtot}) with the $4n\times 4n$ block matrix
\begin{equation*}
B=\begin{pmatrix}
\sqrt \eta\, I&\sqrt{1-\eta}\,I\\
-\sqrt{1-\eta}\,I&\sqrt\eta\, I
\end{pmatrix},
\label{B}
\end{equation*}
and then integrating over the environment variables $\mathbf{x}_{env}$,
we arrive at the output state Wigner function 
\begin{equation}
W_{out}^{\mbox{\boldmath$\alpha$}}(\mathbf{x}_{out})=
\frac1{\sqrt{|V_{out}|}}
\exp\left[-\frac12
\left(\mathbf{x}_{out}-\sqrt{2\eta}\,\mbox{\boldmath$\alpha$}\right)^{\top}
V_{out}^{-1}
\left(\mathbf{x}_{out}-\sqrt{2\eta}\,\mbox{\boldmath$\alpha$}\right)
\right].
\label{Wout}
\end{equation}

Finally, the average output state $\overline{\rho}_{out}=\int d\mbox{\boldmath$\alpha$}\ P( \mbox{\boldmath$\alpha$})
\rho^{\mbox{\boldmath$\alpha$}}_{out}
\equiv \mathcal{L}\left[\overline{\rho}_{in}\right]$, will be characterized by the Wigner function
\begin{equation}
\overline{W}_{out}(\mathbf{x}_{out})=
\int d{\mbox{\boldmath$\alpha$}} P({\mbox{\boldmath$\alpha$}})
W_{out}^{\mbox{\boldmath$\alpha$}}(\mathbf{x}_{out}),
\label{WoutAveraged}
\end{equation}
having covariance matrix
\begin{equation}
\overline{V}_{out}=V_{out}+\eta\, V_{cl}.
\label{Voutave}
\end{equation}


\section{The memory model}
\label{Model}

To introduce the memory effect among different channel uses, we consider the environment modes initially in a correlated state, actually a multimode squeezed vacuum. 
We are going to use multimode squeezed states depending on a single parameter as introduced in Ref.\cite{Lo}, so that we choose the covariance matrix $V_{env}$ as
\begin{equation}
V_{env}=\frac{1}{2}\left[
\begin{array}{cc}
\exp(s\Omega) & 0\\
0 & \exp(-s\Omega)
\end{array}
\right].
\label{Venv}
\end{equation}
Here, the $n\times n$ matrix $\Omega$ is taken to be
\begin{equation}
\Omega=
\begin{pmatrix}
0&1&\hdotsfor{3}&0\\
1&0&1&\hdotsfor{2}&0\\
\vdots&1&0&1&\hdotsfor{1}&0\\
\vdots&\vdots&\ddots&\ddots&\ddots&&\\
\vdots&\vdots&&\ddots&\ddots&1\\
0&0&\hdotsfor{2}&1&0\\
\end{pmatrix},
\label{Om}
\end{equation}
and the parameter $s\in\mathbb{R}$ represents the memory strength (for $s=0$ we recover the memoryless case).
Notice that the memory effect is symmetric among all modes. Moreover, it decays over the number of uses making the model quite realistic. 
Actually, if we consider the set of $V_{env}$ matrix elements belonging to a fixed row as a discrete function, this is well fitted by a Gaussian having 
width $2\sqrt{|s|}$. Such exponential decay of correlations makes
the noise in the channel non-Markovian and the channel not forgetful \cite{Nila}.

Since we would explore the usefulness of entangled inputs in the presence of memory,
we then consider
$\sigma_{in}$ having covariance matrix like that of Eq.(\ref{Venv}), that is
\begin{equation}
V_{in}=\frac{1}{2}\left[
\begin{array}{cc}
\exp(r\Omega) & 0\\
0 & \exp(-r\Omega)
\end{array}
\right],
\label{Vin}
\end{equation}
where $r\in\mathbb{R}$ is the entanglement parameter (for $r=0$ we recover separable vacuum inputs).

To respect the input energy constraint, the added number of photons per mode by entanglement, 
\begin{equation*}
\frac{1}{2n}{\rm Tr}(V_{in})-\frac{1}{2},
\label{add}
\end{equation*}
must necessarily be a fraction of $N$, say $\theta_n N$ with $0\le \theta_n \le 1$.
This limits the range of $r$ to a finite interval of $\mathbb{R}$.
In particular, defining 
\begin{equation*}
\theta_n:=\frac{{\rm Tr}(V_{in})-n}{2nN},
\label{th}
\end{equation*}
 the roots of the equation $ \theta_n=1 $
determine the minimum (negative) and maximum (positive) allowed values of $r$.

Besides entanglement in input, we could introduce classical correlations as well
through $V_{cl}$ [hence $P( \mbox{\boldmath$\alpha$})$]. 
Since entanglement adds a fraction $\theta_n N$ of photons per mode, the diagonal of $V_{cl}$ must
contain a fraction $(1-\theta_n)N$ of photons, while off diagonal could introduce correlations.

We choose the covariance matrix $V_{cl}$ describing classical noise to be of the same form of $V_{env}$ and $V_{in}$, that is 
\begin{equation}
V_{cl}:=\frac{2nN(1-\theta_n)}{{\rm Tr}(Y)}\, Y,
\label{Vcl}
\end{equation}
where 
\begin{equation*}
Y=\frac{1}{2}\left[
\begin{array}{cc}
\exp(y\Omega) & 0\\
0 & \exp(-y\Omega)
\end{array}
\right],
\label{Y}
\end{equation*}
with the parameter $y\in\mathbb{R}$ describing classical correlations.

It is worth remarking that the choice of the form  of covariance matrices $V_{env}$, $V_{in}$, and $V_{cl}$ allows us to diagonalise them 
 in the same basis. The eigenvalues of this kind of matrices are shown 
in Appendix A. Moreover, in Appendix B we show that the amount of average photon number per mode remains finite in these matrices by increasing their dimension.


\section{The Holevo-$\chi$ quantity}

Recall that the von Neumann entropy of a Gaussian state $\rho$ with covariance matrix $V$, can be simply written in terms of symplectic eigenvalues of $V$ \cite{Hol}. This means that
\begin{equation*}
S(\rho)=\sum_{k=1}^n
g\left( \left| \nu^{(V)}_{k}\right| -
\frac{1}{2}\right),
\label{S}
\end{equation*}
 where
 \begin{equation*}
g\left( x\right):=
\left( x+1\right) \log_2 \left( x+1\right) -x\log_2 x,
\label{gx}
\end{equation*} 
and $\nu^{(V)}_k$ are the symplectic eigenvalues of $V$, i.e. $\pm i\nu^{(V)}_k$ are the eigenvalues of $\Sigma^{-1}V$ with $\Sigma$ the symplectic $2n\times 2n$ block matrix
\begin{equation*}
\Sigma:=\left(
\begin{array}{cc}
0 & I  \\
-I & 0 
\end{array}
\right).
\end{equation*}

The classical capacity of a memoryless quantum channel is determined by means of the output 
Holevo-$\chi$ quantity
 \begin{equation}
\chi_n:=S\left(\overline{\rho}_{out}\right)-\int d{\mbox{\boldmath$\alpha$}}
P({\mbox{\boldmath$\alpha$}}) S\left(\rho_{out}^{\mbox{\boldmath$\alpha$}}\right).
\label{chi}
\end{equation}
In case of  memoryless quantum channels, this quantity represents the highest achievable rate for a fixed encoding scheme. Thus, its maximization overall possible encodings (i.e. possible input states and probability distributions-Gaussians for the conjecture of Ref.\cite{Hol}) gives the classical capacity.

Assuming that $\chi$ plays the same role for memory channels as well and assuming the possibility to
extend the conjecture of Ref.\cite{Hol}, we should maximize Eq.\eqref{chi} overall Gaussian states
$\sigma_{in}$ and distributions $P({\mbox{\boldmath$\alpha$}})$.  As a first step, here we consider the
maximization overall Gaussian inputs of the form of Eqs.(\ref{Vin}) and (\ref{Vcl}).

Since the covariance matrices  (\ref{Vout}), (\ref{Voutave})  do not depend on ${\mbox{\boldmath$\alpha$}}$, we get $\chi$ simply depending on the two parameters $r,y$
\begin{equation}
\chi_n(r,y)=\sum_{k=1}^{n}\left[ g\left( \left| \nu_{k}^{(\overline{V}^{out})}\right| -
\frac{1}{2}\right) -
g\left( \left| \nu_{k}^{(V^{out})}\right| -
\frac{1}{2}\right)
\right], 
\label{chiry}
\end{equation}
with $\nu _{k}^{(\overline{V}^{out})}$, $\nu _{k}^{(V^{out})}$ the symplectic eigenvalues of
$\overline{V}_{out}$ and $V_{out}$ respectively.

By applying the model of Section \ref{Model} we get the following result for the symplectic eigenvalues (see Appendix A):
\begin{eqnarray}
\nu^{(V^{out})}_{k}&=&
\frac12 \sqrt{\eta^2+(1-\eta)^2+2\eta\,(1-\eta)\cosh\varphi_{sr}^{(n)}},
\label{simVo}\\
\nu^{(\overline{V}^{out})}_{k}&=&
\frac12\sqrt{\eta^2+(1-\eta)^2+\eta^2K_n^2+
2\eta\,(1-\eta)\cosh\varphi^{(n)}_{sr}+2\eta^2 K_n\cosh\varphi^{(n)}_{ry}+2\eta(1-\eta)K_n\cosh\varphi^{(n)}_{sy}},
\label{simVoAv}
\end{eqnarray}
where
\begin{equation*}
\varphi^{(n)}_{lm}:=2(l-m)\cos\left(\frac{\pi k}{n+1}\right),\,\,\,\,l,m=r,y,s,
\label{varphi}
\end{equation*}
\begin{equation}
K_n=\frac{2nN(1-\theta_n)}{\sum_{k=1}^n\cosh\left(2y\cos\left(\frac{\pi k}{n+1}\right)\right)},
\label{Kappa}
\end{equation}
\begin{equation}
\theta_n=
\frac{\sum_{k=1}^n\cosh\left(2r\cos\left(\frac{\pi k}{n+1}\right)\right)-n}{2nN}.
\label{theta}
\end{equation}
From a geometrical point of view, each symplectic eigenvalue of Eq.(\ref{simVo}) can be interpreted
as the relation among triangle sides of length $\eta$, $1-\eta$ and $2\nu^{(V^{out})}_{k}$ with the
angle between sides of length $\eta$ and $1-\eta$ equal to $\pi-i\varphi_{sr}^{(n)}$.
Analagously, each symplectic eigenvalue of Eq.(\ref{simVoAv}) can be interpreted as the relation among 
tetrahedron faces of area $\eta$, $1-\eta$, $\eta K_n$ and $2\nu^{(\overline{V}^{out})}_{k}$
with dihedral angles $i\varphi_{sr}^{(n)}$,  $i\varphi_{ry}^{(n)}$ and $i\varphi_{sy}^{(n)}$ among the first three faces.
\begin{figure}
\centerline{\includegraphics{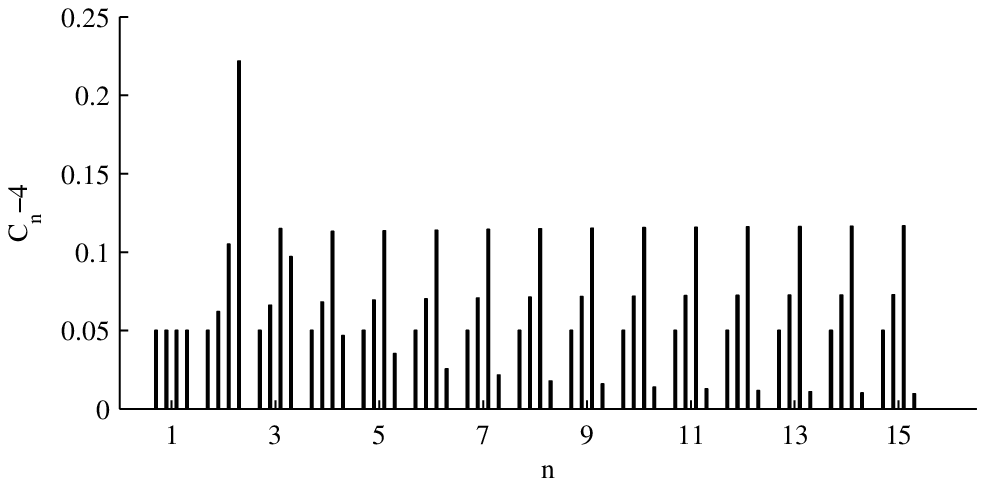}\qquad\includegraphics{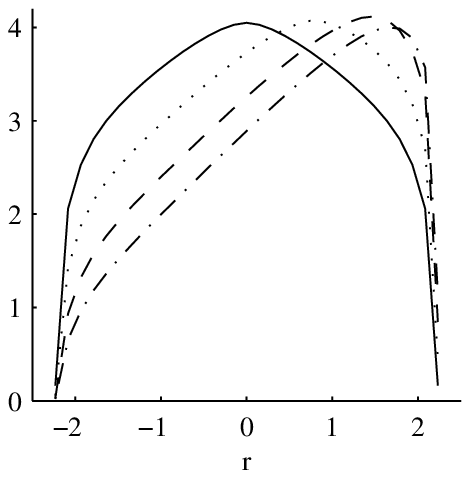}}
\caption{
On the left, the quantity $C_n$  is plotted versus $n$. For each $n$, bars from 
left to right refer to $s=0$, $0.8$, $1.6$, $2.5$ respectively. The other parameters are $N=8$ and $\eta=0.7$. 
On the right the quantity
$\lim_{n\rightarrow\infty}\left[\max_{y\in\mathbb{R}}\frac1n\chi_n(r,y)\right]$
is plotted versus $r$.
Solid, dotted, dashed and dashed-dotted curves refer to $s=0$, $0.8$, $1.6$, $2.5$ respectively.
The curves attain maxima at $4.05$, $4.07$, $4.12$, $4$ respectively.
The other parameters are $N=8$ and $\eta=0.7$.
}
\label{HolevoFig}
\end{figure}

We can now define the quantity
\beq
C_n:=\max_{r,y\in\mathbb{R}}\;\frac{1}{n}\chi_n(r,y).
\label{Cn}
\eeq
which would be a marker of the behavior of the capacity for any number $n$ of channel uses.
Substituting Eqs.\eqref{simVo} and~\eqref{simVoAv} in the relation~\eqref{chiry} and maximizing it over
possible values of variables $r,y$ we can evaluate the quantity~\eqref{Cn}.
Such a quantity is plotted in Fig.\ref{HolevoFig}-left versus $n$ for different values of $s$.
On the one hand, for $n=1$ (or equivalently for $s=0$)  we recover the value $g(\eta N)$ of the memoryless channel capacity \cite{GG}.
On the other hand, it is clear the possibility to overcome the memoryless channel capacity (by using entangled inputs)  when $s\neq 0$. In such cases we may notice different transient behavior by increasing $n$.

By taking the limit $n\rightarrow\infty$ of $C_n$  we find the asymptotic behavior.
Actually, this limit results the limit of Riemann
sums thus leading to an integral (as described in Appendix B).
Due to the regularity of the function in Eq.\eqref{chiry} we can write
\begin{equation}
C:=\lim_{n\to \infty}C_n=\max_{r,y\in\mathbb{R}}\;\frac{1}{\pi}\int_0^\pi\left[
g\left( \left| \nu^{(\overline{V}^{out})}\right| - \frac{1}{2}\right) - 
g\left( \left| \nu^{(V^{out})}\right| - \frac{1}{2}\right)
\right]d\xi,
\label{hol_cap}
\end{equation}
where $\nu^{(V^{out})}$ and  $\nu^{(\overline{V}^{out})}$ are given by Eqs.~\eqref{simVo}
and~\eqref{simVoAv} with $K_n$ replaced by $K$ (see Appendix~B) and $\varphi^{(n)}_{lm}$ replaced by $\varphi_{lm}:=2\,(l-m)\cos\xi$. 

In Fig.\ref{HolevoFig}-right the quantity
$\lim_{n\to\infty}\left[\max_{y\in\mathbb{R}}\frac1n\chi_n(r,y)\right]$
is plotted versus $r$. The maxima correspond to values of the quantity \eqref{hol_cap}.
For $s\neq 0$, it clearly shows the possibility to overcome the memoryless channel capacity by using
entangled inputs ($r\neq 0$).


\section{Conventional Decodings}

We are now going to consider information transmission rates by conventional decoding procedures like heterodyne and homodyne measurements \cite{Bra,GMD}.

\subsection{Decoding by joint quadratures measurements}

Let us consider decoding at the output by joint measurement of canonical variables on each mode
(heterodyne measurement). This is described by the probability operator measure (POM)
$\otimes_{j=1}^n\left(|\zeta_j\rangle\langle\zeta_j|/\pi\right)$ with $|\zeta_j\rangle$ coherent state of
the $j$-th mode having complex amplitude $\zeta_j$.

In analogy with the vector $\mbox{\boldmath$\alpha$}$  we are going to introduce the vector
$\mbox{\boldmath$\zeta$}=(\Re(\zeta_1),\dots,\Re(\zeta_n),\Im(\zeta_1),\dots,\Im(\zeta_n))^{\top}$. 
The probability for coherent amplitudes $\mbox{\boldmath$\zeta$}$ is related to the
probability for the corresponding quadratures $\mathbf{x}_{out}$, which is given by the Wigner
function~\eqref{Wout}. Thus, the conditional probability of getting
$\mbox{\boldmath$\zeta$}$ at the output given the encoded $\mbox{\boldmath$\alpha$}$ at input results
\begin{equation}
P(\mbox{\boldmath$\zeta$}|\mbox{\boldmath$\alpha$})=
\frac1{\pi^{n}\sqrt{|V_{\mbox{\tiny\boldmath$\zeta$}|\mbox{\tiny\boldmath$\alpha$}}|}}
\exp\left[-
\left(\mbox{\boldmath$\zeta$}-\sqrt{\eta}\,\mbox{\boldmath$\alpha$}\right)^{\top}
V_{\mbox{\tiny\boldmath$\zeta$}|\mbox{\tiny\boldmath$\alpha$}}^{-1}
\left(\mbox{\boldmath$\zeta$}-\sqrt{\eta}\,\mbox{\boldmath$\alpha$}\right)
\right],
\label{Pcond1}
\end{equation}
where
\begin{equation*}
V_{\mbox{\tiny\boldmath$\zeta$}|\mbox{\tiny\boldmath$\alpha$}}:=V_{out}+\frac12,
\end{equation*}
is the relation for quadratures' covariance matrices.
Here, the term $1/2$ represents the quadrature vacuum noise added by heterodyne measurement. 
Then,  using Eqs.\eqref{WoutAveraged}
and \eqref{Voutave} we get the output probability
\begin{equation}
P(\mbox{\boldmath$\zeta$})=
\frac1{\pi^n\sqrt{|V_{\mbox{\tiny\boldmath$\zeta$}}|}}
\exp\left[-\mbox{\boldmath$\zeta$}^{\top}
V_{\mbox{\tiny\boldmath$\zeta$}}^{-1}
\mbox{\boldmath$\zeta$}
\right],
\label{Pout}
\end{equation}
where
\begin{equation*}
V_{\mbox{\tiny\boldmath$\zeta$}}:=\overline{V}_{out}+\frac12.
\label{Vz}
\end{equation*}
\begin{figure}
\centerline{\includegraphics{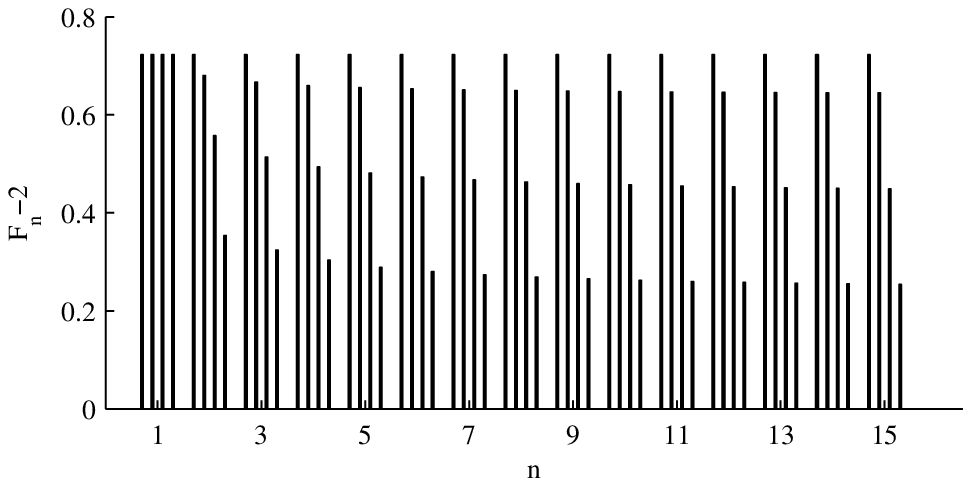}\qquad\includegraphics{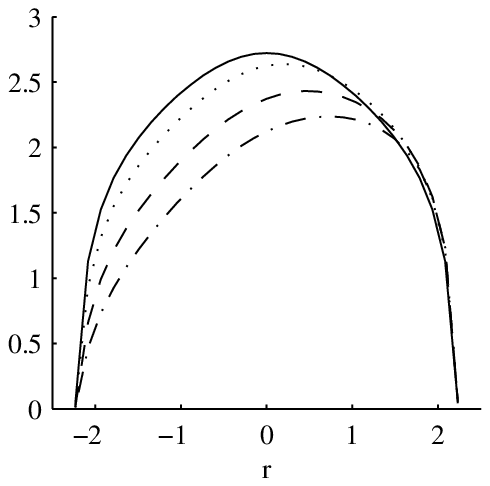}}
\caption{
On the left, the quantity $F_n$ (heterodyne case) is plotted versus $n$. For each $n$, bars from 
left to right refer to $s=0$, $0.8$, $1.6$, $2.5$ respectively. The other parameters are $N=8$ and $\eta=0.7$. 
On the right the quantity
$\lim_{n\to\infty}\left[\max_{y\in\mathbb{R}}\frac1n I(\mathbf{Z}:\mathbf{A})(r,y)\right]$ 
is plotted versus $r$.
Solid, dotted, dashed and dashed-dotted curves refer to $s=0$, $0.8$, $1.6$, $2.5$ respectively.
The curves attain maxima at $2.72$, $2.64$, $2.43$, $2.24$ respectively.
The other parameters are $N=8$ and $\eta=0.7$.}
\label{HeterodyneFig}
\end{figure}

The Shannon differential entropy for a stochastic variable $\Phi$ taking real values $\phi$ distributed with probability $P(\phi)$ reads
\begin{equation}
H({\Phi})=-\int_{supp\,\Phi} P(\phi)\log_2P(\phi)\,d\phi.
\label{ShannonDef}
\end{equation}
Applying this definition to Eqs.\eqref{Pal}, \eqref{Pcond1} and \eqref{Pout}   we arrive at the mutual
information~\footnote{This expression has been obtained by exploiting the commutativity of covariance
matrices.}
\beqa
I(\mbox{\boldmath$Z$}:\mbox{\boldmath$A$}) 
&=&H({\mbox{\boldmath$Z$}})-H({\mbox{\boldmath$Z$}|\mbox{\boldmath$A$}})\nonumber\\
&=&\frac12\log_2\left|\left(\overline V_{out}+\frac12\right)\left(V_{out}+\frac12\right)^{-1}\right|.
\label{IZA}
\eeqa

By using the model of Section \ref{Model} we explicitly get
\begin{equation}
I(\mbox{\boldmath$Z$}:\mbox{\boldmath$A$})=
\frac12\sum_{k=1}^n\,\left[\log_2\left(F_{+,k}\right)+\log_2\left(F_{-,k}\right)\right],
\label{HDi}
\end{equation}
where
\begin{equation*}
F_{\pm,k}=
1+K_n\frac{\eta\,e^{\pm2y\cos\left(\frac{\pi k}{n+1}\right)}}
{\eta\,e^{{\pm2r\cos\left(\frac{\pi k}{n+1}\right)}}+(1-\eta)\,e^{{\pm2s\cos\left(\frac{\pi k}{n+1}\right)}}+1}.
\end{equation*}

The quantity $F_n:=\max_{r,y\in\mathbb{R}}I(\mbox{\boldmath$Z$}:\mbox{\boldmath$A$})/n$ is plotted in Fig.\ref{HeterodyneFig}-left versus $n$ for different values of $s$.
On the one hand, for $n=1$ (or equivalently for $s=0$)  we recover the value $\log_2(1+\eta N)$ of the memoryless channel rate \cite{GG}.
On the other hand, it is clear that in this case the memory effects worsen the rate with respect to the memoryless case.

By taking the limit $n\rightarrow\infty$ of $F_n$  we find the asymptotic behavior of  the rate.
Actually, this limit results the limit of Riemann
sums thus leading to an integral (as described in Appendix B).
Due to the regularity of the function in Eq.\eqref{HDi} we can write
\begin{equation}
F:=\lim_{n\to\infty}F_n=\max_{r,y\in\mathbb{R}}\;\frac{1}{\pi}\int_0^\pi\log_2\left(
1+K\frac{\eta\,e^{2y\cos\xi}}{\eta\,e^{2r\cos\xi}+(1-\eta)\,e^{2s\cos\xi}+1}
\right)d\xi.
\label{HDrates}
\end{equation}
In Fig.\ref{HeterodyneFig}-right the quantity $\lim_{n\to\infty}\left[\max_{y\in\mathbb{R}}\frac1n
I(\mathbf{Z}:\mathbf{A})(r,y)\right]$ is plotted versus $r$ for different values of $s$. The attained
maxima correspond to values of $F$. We see that the maximum for $s=0$ implies $r=0$ (and also $y=0$)
[this is a global maximum for the mutual information]. Nevertheless, for $s\neq 0$, the maxima are
obtained for $r\neq 0$. Hence, entangled inputs give a higher rate with respect to separable ones in
presence of memory, although such a rate is always smaller than the one for memoryless case.


\subsection{Decoding by single quadratures measurements}

Let us consider decoding at the output by measurement of a single canonical variable on each mode. This
is described, e.g. by the POM $\otimes_{j=1}^n\left(|\Re(\zeta_j)\rangle\langle \Re(\zeta_j)|\right)$. 
In the following it will be useful to write
$\mbox{\boldmath$\alpha$}=(\mbox{\boldmath$\alpha$}_R,\mbox{\boldmath$\alpha$}_I)$, and
$\mbox{\boldmath$\zeta$}=(\mbox{\boldmath$\zeta$}_R,\mbox{\boldmath$\zeta$}_I)$.
Then, we have to consider information only encoded in $\mbox{\boldmath$\alpha$}_R$.
Hence, while keeping the usual energy constrain, we will consider the matrix $Y$ in Eq.\eqref{Vcl} now written as
\begin{equation*}
Y=\left[
\begin{array}{cc}
e^{y\Omega} & 0\\
0 & 0
\end{array}
\right].
\end{equation*} 
Interpreting the Wigner function~\eqref{Wout} as the probability for
$\mbox{\boldmath$\zeta$}$ conditioned to $\mbox{\boldmath$\alpha$}$, 
and taking into account that the stochastic variables  $\mbox{\boldmath$\alpha$}_R$, $\mbox{\boldmath$\zeta$}_R$ are independent from $\mbox{\boldmath$\alpha$}_I$, $\mbox{\boldmath$\zeta$}_I$,
we get, by integrating Eq.\eqref{Wout} over $\mbox{\boldmath$\zeta$}_I$,
\begin{equation}
P(\mbox{\boldmath$\zeta$}_R|\mbox{\boldmath$\alpha$}_R)
=\frac1{\pi^{n/2}\sqrt{\left|V_{out}^{(11)}\right|}}\exp\left[
-\left(\mbox{\boldmath$\zeta$}_R-\sqrt{\eta}\,\mbox{\boldmath$\alpha$}_R\right)^{T}
\left(V_{out}^{(11)}\right)^{-1}
\left(\mbox{\boldmath$\zeta$}_R-\sqrt{\eta}\,\mbox{\boldmath$\alpha$}_R\right)
\right].
\label{Pcondhom}
\end{equation}
Analogously we can consider the Wigner function \eqref{WoutAveraged} as 
the probability for
$\mbox{\boldmath$\zeta$}$ and by integrating it over $\mbox{\boldmath$\zeta$}_I$ we get
\begin{equation}
P(\mbox{\boldmath$\zeta$}_R)
=\frac1{\pi^{n/2}\sqrt{\left|\overline V_{out}^{(11)}\right|}}\exp\left[
-\mbox{\boldmath$\zeta$}_R^{T}
\left(\overline V_{out}^{(11)}\right)^{-1}
\mbox{\boldmath$\zeta$}_R
\right].
\label{Pouthom}
\end{equation}
Finally, using the probabilities \eqref{Pcondhom} and \eqref{Pouthom}
in Eq.\eqref{ShannonDef}, we arrive at 
\begin{eqnarray*}
I(\Re\mbox{\boldmath$Z$}:\Re\mbox{\boldmath$A$}) 
&=&H(\Re\mbox{\boldmath$Z$})-H(\Re\mbox{\boldmath$Z$}|\Re\mbox{\boldmath$A$})\nonumber\\
&=&\frac12\log_2\left|\left(\overline V_{out}^{(11)}\right)\left(V_{out}^{(11)}\right)^{-1}\right|.
\end{eqnarray*}
By using the model of Section \ref{Model} we explicitly get
\begin{equation}
I(\Re\mbox{\boldmath$Z$}:\Re\mbox{\boldmath$A$})=
\frac12\sum_{k=1}^n\log_2\left(
1+2K_n\frac{\eta\,e^{2y\cos\left(\frac{\pi k}{n+1}\right)}}
{\eta\,e^{{2r\cos\left(\frac{\pi k}{n+1}\right)}}+(1-\eta)\,e^{{2s\cos\left(\frac{\pi k}{n+1}\right)}}}
\right).
\label{IRezRea}
\end{equation}

\begin{figure}
\centerline{\includegraphics{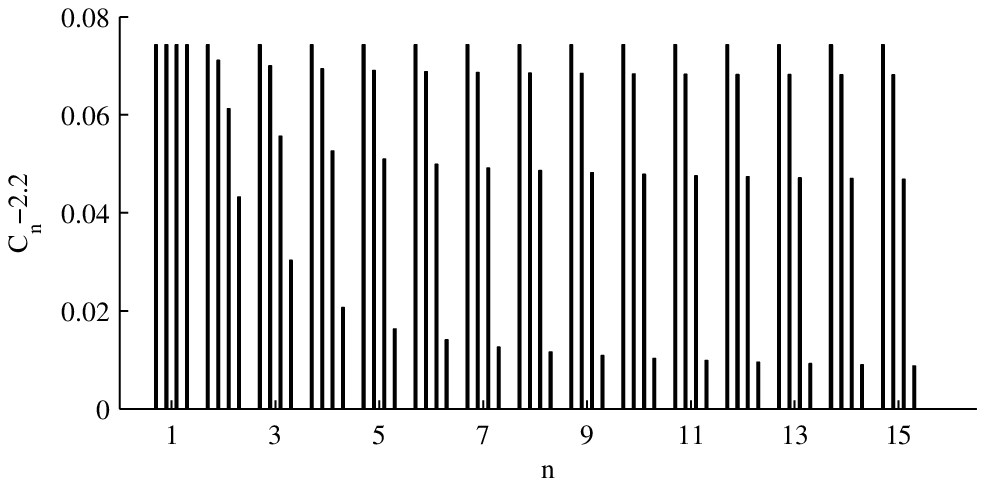}\qquad\includegraphics{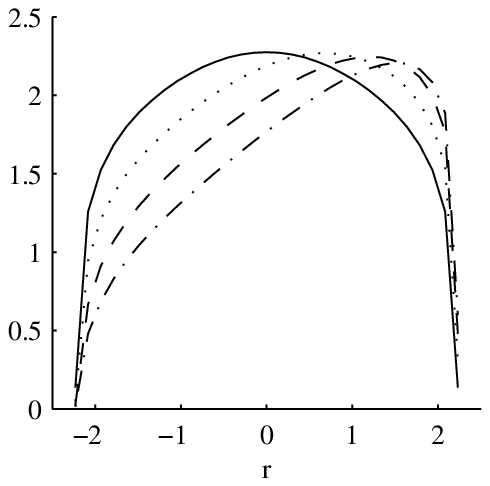}}
\caption{
On the left, the quantity $F_n$ (homodyne case) is plotted versus $n$. For each $n$, bars from 
left to right refer to $s=0$, $0.8$, $1.6$, $2.5$ respectively. The other parameters are $N=8$ and $\eta=0.7$. 
On the right the quantity
$\lim_{n\to\infty}\left[\max_{y\in\mathbb{R}}\frac1n I(\Re\mathbf{Z}:\Re\mathbf{A})(r,y)\right]$ 
is plotted versus $r$.
Solid, dotted, dashed and dashed-dotted curves refer to $s=0$, $0.8$, $1.6$, $2.5$ respectively.
The curves attain maxima at $2.274$, $2.27$, $2.25$, $2.21$ respectively.
The other parameters are $N=8$ and $\eta=0.7$.}
\label{ujos2}
\end{figure}

The quantity $F_n:=\max_{r,y\in\mathbb{R}}I(\Re\mbox{\boldmath$Z$}:\Re\mbox{\boldmath$A$})/n$ is plotted
in Fig.\ref{ujos2}-left versus $n$ for different values of $s$.  On the one hand, for $n=1$ (or
equivalently for $s=0$)  we recover the value $(1/2)\log_2(1+4\eta N)$ of the memoryless channel
rate~\cite{GG}.  On the other hand, it is clear that in this case the memory effects worsen the rate with
respect to the memoryless case.

By taking the limit $n\rightarrow\infty$ of $F_n$  we find the asymptotic behavior of  the rate.
Actually, this limit results the limit of Riemann
sums thus leading to an integral (as described in Appendix B).
Due to the regularity of the function in Eq.\eqref{IRezRea} we can write
\begin{equation}
F:=\lim_{n\to\infty}F_n=\max_{r,y\in\mathbb{R}}\;\frac{1}{2\pi}\int_0^\pi\log_2\left(
1+2K\frac{\eta\,e^{2y\cos\xi}}{\eta\,e^{2r\cos\xi}+(1-\eta)\,e^{2s\cos\xi}}
\right)d\xi.
\label{SQcase}
\end{equation}

In Fig.\ref{ujos2}-right the quantity $\lim_{n\to\infty}\left[\max_{y\in\mathbb{R}}\frac1n
I(\Re\mathbf{Z}:\Re\mathbf{A})(r,y)\right]$ is plotted versus $r$ for different values of $s$. The
attained maxima correspond to values of $F$. We see that the maximum for $s=0$ implies $r=0$ (and also
$y=0$) [this is a global maximum for the mutual information as in heterodyne case]. Nevertheless, for
$s\neq 0$, the maxima are obtained for $r\neq 0$. Hence, entangled inputs give a higher rate with respect
to separable ones in presence of memory, although such rate is always smaller than the rate of heterodyne
detection (for the same values of $s$).


\section{Conclusion}

We have presented a realistic model to describe memory effects in a lossy bosonic channel.  These effects
have nontrivial long range correlations (non-Markovian). Notwithstanding, the model has allowed us to
characterize the channel over an arbitrary number of uses for classical information transmission.
Therefore, the asymptotic behavior of a Gaussian memory channel has been studied for the first time.

We have shown the usefulness of entangled inputs in the presence of memory. In particular, we have shown
that they can enhance the classical capacity with respect to the memoryless case. This has been done by
maximizing the Holevo-$\chi$ quantity over a class of Gaussian states.  We have also shown the utility of
entangled inputs for rates achievable by conventional decoding procedures (heterodyne and homodyne
measurements), although these do not exceed the classical capacity of the memoryless channel.  
\begin{figure}[t]
\centerline{\includegraphics{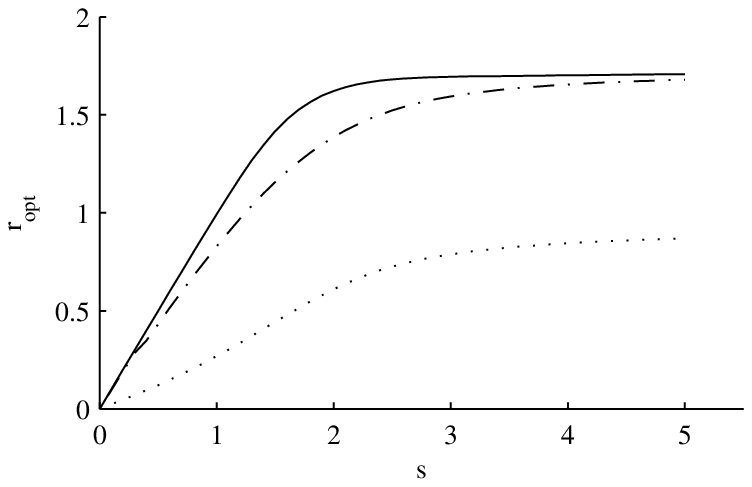}\qquad\includegraphics{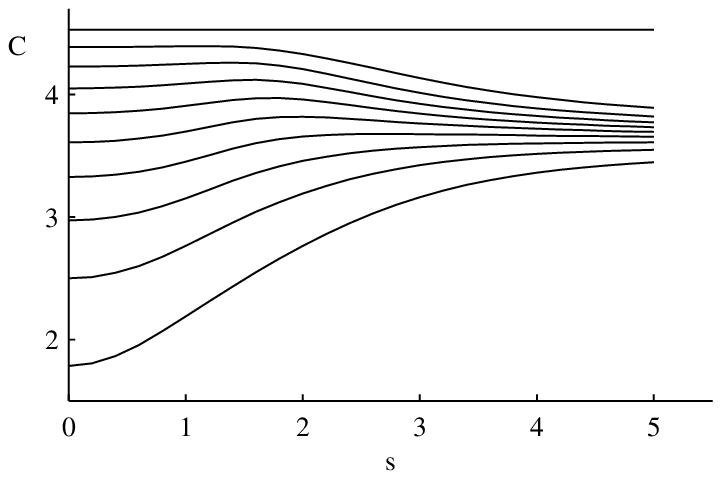}}
\caption{
On the left, the optimal $r$ value for the quantities of Eqs.\eqref{hol_cap} (solid line), \eqref{HDrates} (dotted line) and \eqref{SQcase} (dashed-dotted line) is shown versus $s$.
The value of the other parameters are $N=8,\eta=0.7$.
On the right, the quantity $C$ of Eq.\eqref{hol_cap} is plotted versus $s$ for values of $\eta$ going
from $0.1$ (bottom curve) to $1$ (top curve) with step $0.1$. The value of the other parameter is $N=8$.}
\label{concl}
\end{figure}

The usefulness of entangled inputs is summarized in Fig.\ref{concl}-left where we can see that the
optimal entanglement degree smoothly varies with the degree of memory. The absence of any kink excludes
phenomena similar to phase transitions in contrast to what happen for qubit memory channels \cite{VP}.
Still, in Fig.\ref{concl}-right we can see the enhancement of the classical capacity in the presence of
memory for channels characterized by different values of transmittivity $\eta$.  
Such enhancement grows thinner and thinner as $\eta$ approaches one. Actually, 
the maximum in each curve shifts towards $s=0$ as $\eta$ approaches one.
For $\eta=1$ the quantity $C$ does not longer depend on $s$ because the environment no longer affects inputs.
We also notice that for increasing values of $s$, curves corresponding to different values of $\eta$ (except $\eta=1$) flow together towards a single value. This is presumably due to the fact that
for large values of $s$ the class of Gaussian states used for the maximization is too small to get the optimal solution (which seems in agreement with the saturation of the optimal $r$ value).

In conclusion, we believe that this study paves the way for a deeper characterization of lossy bosonic
memory channels.  The next step would be the full maximization of the $\chi$ quantity overall Gaussian
inputs.  It remains an open question what could be the optimal decoding procedure able to attain a
capacity higher than the memoryless one.


\acknowledgments

O. P. thanks Yu. N. Maltsev for  fruitful discussions.

\section*{APPENDIX A}

Let us consider the $n\times n$ matrix $\Omega$ of Eq.(\ref{Om}).
To obtain its eigenvalues we first notice that $\Omega=2I-T$, where $T$ is the following $n\times n$ matrix
\begin{equation*}
T:=
\begin{pmatrix}
2&-1&0&\hdotsfor{3}&0\\
-1&2&-1&0&\hdotsfor{2}&0\\
0&-1&2&-1&0&\hdotsfor{1}&0\\
\vdots&\ddots&\ddots&\ddots&\ddots&\ddots&\vdots\\
0&\hdotsfor{1}&0&-1&2&-1&0\\
0&\hdotsfor{2}&0&-1&2&-1\\
0&\hdotsfor{3}&0&-1&2\\
\end{pmatrix}.
\end{equation*}
Being a finite-difference counterpart of the differential operator $-\frac{d^2}{dx^2}$, the matrix $T$ plays an important role in computational mathematics
and its properties are well known. 
For instance, it can be shown that it
 has the following eigenvalues ~\cite{Demmel}:
\begin{equation*}
\lambda_j^{(T)}=2\left(1-\cos\frac{\pi{}j}{n+1}\right),
\end{equation*}
with corresponding normalized eigenvectors with components
\begin{equation*}
v_{j,k}^{(T)}=\sqrt\frac2{n+1}\sin\frac{jk\pi}{n+1},\quad k=1,\ldots,n.
\end{equation*}
Thus, the matrix $\Omega$ will have the same eigenvectors $v^{(T)}_j$ and its eigenvalues turn out to be
\begin{equation}
\lambda_j^{(\Omega)}=2-\lambda^{(T)}_j=2\cos\frac{\pi{}j}{n+1}.
\label{omega_eval}
\end{equation}

For any matrix like
\begin{equation}
A = \frac12
\begin{pmatrix}
e^{\gamma\Omega}&0\\
0&e^{-\gamma\Omega}
\end{pmatrix},\quad \gamma\in\mathbb{R},
\label{Amatrix}
\end{equation}
we get the eigenvalues 
\begin{equation}
\lambda^{(A)}_{\pm,k}=\frac12e^{\pm2\gamma\cos\left(\frac{\pi k}{n+1}\right)},
\label{Alyambda}
\end{equation}
as direct consequence of Eq.\eqref{omega_eval}.

Since $V_{env}$, $V_{in}$ and $V_{cl}$ have the same block structure in terms of $\Omega$, they
can be diagonalised in the same basis. Hence, by taking into account Eqs.\eqref{Vout}, \eqref{Voutave}, we straightforwardly get the eigenvalues of  $V_{out}$ and $\overline{V}_{out}$
\begin{eqnarray}
\lambda^{(V_{out})}_{\pm,k}&=&\frac12\left(\
\eta e^{\pm2r\cos\left(\frac{\pi k}{n+1}\right)}+
(1-\eta) e^{\pm2s\cos\left(\frac{\pi k}{n+1}\right)}
\right),\nonumber\\
\lambda^{(\overline V_{out})}_{\pm,k}&=&\lambda^{(V_{out})}_{\pm,k}+
\frac12e^{\pm2y\cos\left(\frac{\pi k}{n+1}\right)}\eta K_n.\nonumber
\end{eqnarray}
with $K_n$ given by Eq.\eqref{Kappa}.

Finally, by definition of symplectic eigenvalues we get
\begin{eqnarray}
\nu^{(V_{out})}_{k}&=&\sqrt{\lambda^{(V_{out})}_{+,k}\lambda^{(V_{out})}_{-,k}},\nonumber\\
\nu^{(\overline{V}_{out})}_{k}&=&\sqrt{\lambda^{(\overline{V}_{out})}_{+,k}\lambda^{(\overline{V}_{out})}_{-,k}},\nonumber
\end{eqnarray}
leading to Eqs.\eqref{simVo} and \eqref{simVoAv}.


\section*{APPENDIX B}

We show that dealing with matrices of the form of Eq.\eqref{Amatrix}, the average number of photons per mode remains finite even in the limit $n\to\infty$. To this end we consider
\begin{equation}
\lim_{n\rightarrow\infty}\frac{\Tr(A)}{2n}=
\lim_{n\rightarrow\infty}\frac{\Tr(A)}{2(n+1)}=\frac14\left(
\lim_{n\rightarrow\infty}\frac{\Tr(e^{\gamma\Omega})}{n+1}+\lim_{n\rightarrow\infty}\frac{\Tr(e^{-\gamma\Omega})}{n+1}
\right).
\label{limits}
\end{equation}
By taking into account Eq.\eqref{Alyambda} we can rewrite the first term at right hand side of 
Eq.\eqref{limits} as 
\begin{equation*}
\lim_{n\rightarrow\infty}\frac{\Tr(e^{\gamma\Omega})}{n+1}=
\lim_{n\rightarrow\infty}\frac{\sum_{k=1}^ne^{2\gamma\cos\frac{\pi k}{n+1}}}{n+1}.
\end{equation*}
This relation is the limit of Riemann sum becoming an integral for the function $f(x)=e^{2\gamma\cos\pi x}$. This leads to a modified
Bessel function of the first kind and zero-order~\cite{Abramowitz}
\begin{equation*}
\lim_{n\rightarrow\infty}\frac{\Tr(e^{\gamma\Omega})}{n+1}=\int_0^1e^{2\gamma\cos\pi x}\,dx=
\frac1\pi\int_0^\pi e^{2\gamma\cos\xi}\,d\xi=I_0(2\gamma).
\end{equation*}
Since the Bessel finction $I_0$ is even, the result of Eq.\eqref{limits} is $I_0(2\gamma)/2$ whose
asymptotic behavior is $e^{2\gamma}/(4\sqrt{\pi\gamma})$ for large $\gamma$.  The existence of a finite
limit in Eq.\eqref{limits} allows us to conclude that the considered matrices (like~\eqref{Amatrix}) give
rise to a physical model.

Finally, by applying the above method to calculate the limit values of $\theta_n$ and $K_n$ given by
relations~\eqref{theta} and~\eqref{Kappa}, we get
\begin{equation}
\begin{split}
&\,\,\theta:=\lim_{n\rightarrow\infty}\theta_n\,\,=\frac{I_0(2r)-1}{2N},\\
&K:=\lim_{n\rightarrow\infty}K_n=\frac{2(N+1/2)-I_0(2r)}{I_0(2y)}.
\end{split}
\label{result_ka}
\end{equation}


\end{document}